\documentclass[conference]{IEEEtran}
\usepackage{hyperref}
\hypersetup{
 unicode=true,
 pdfstartview={FitH},
 colorlinks=true,
 citecolor=blue,
 linkcolor=blue,
 urlcolor=blue
}

\usepackage[linesnumbered,algoruled]{algorithm2e}
\usepackage{algpseudocode}

\usepackage{caption}
\usepackage{subcaption}
\usepackage{multirow}
\usepackage{makecell}
\usepackage[inline]{enumitem}
\usepackage{amssymb}

\usepackage{listings}
\usepackage[comments]{annotations}

\newcommand{\nop}[1]{}

\newcommand{\Com}[1]{}

\usepackage{listings}

\crefname{algocf}{alg.}{algs.}
\Crefname{algocf}{Algorithm}{Algorithms}
\SetKwRepeat{Do}{do}{while}

\usepackage{enumitem}
\usepackage{acronym}
\acrodef{IP}[IP]{intellectual property block}
\acrodef{SoC}[SoC]{System-on-Chip}
\acrodef{IC}[IC]{integrated circuit}
\acrodef{eFPGA}[eFPGA]{embedded field programmable gate array}
\acrodef{FPGA}[FPGA]{field programmable gate array}
\acrodef{RTL}[RTL]{register transfer level}
\acrodef{CLB}[CLB]{configurable logic block}
\acrodef{LUT}[LUT]{look-up table}
\acrodef{HLS}[HLS]{high-level synthesis}
\acrodef{EDA}[EDA]{electronic design automation}
\acrodef{FF}[FF]{flip-flop}
\acrodef{DIP}[DIP]{distinguishing input pattern}
\acrodef{PPA}[PPA]{power, performance, and area}
\acrodef{BLE}[BLE]{basic logic element}
\acrodef{CB}[CB]{connection block}
\acrodef{SB}[SB]{switch block}
\usepackage{booktabs}
\newcolumntype{L}[1]{>{\raggedright\let\newline\\\arraybackslash\hspace{0pt}}m{#1}}
\newcolumntype{C}[1]{>{\centering\let\newline\\\arraybackslash\hspace{0pt}}m{#1}}
\newcolumntype{R}[1]{>{\raggedleft\let\newline\\\arraybackslash\hspace{0pt}}m{#1}}

\usepackage{amsmath,amsfonts}
\usepackage{url}

\begin{document}

\IEEEoverridecommandlockouts
\IEEEpubid{\makebox[\columnwidth]{ 979-8-3503-7608-1/24\$31.00 \copyright2024 IEEE \hfill} \hspace{\columnsep}\makebox[\columnwidth]{ }}

\title{C2HLSC: Can LLMs Bridge the Software-to-Hardware Design Gap?}
\author{Luca Collini~\IEEEmembership{Graduate~Student~Member,~IEEE}, Siddharth Garg~\IEEEmembership{Member,~IEEE}, Ramesh Karri~\IEEEmembership{Fellow,~IEEE}
}

\maketitle
\begin{abstract}
 High Level Synthesis (HLS) tools offer rapid hardware design from C code, but their compatibility is limited by code constructs. This paper investigates Large Language Models (LLMs) for refactoring C code into HLS-compatible formats. We present several case studies by using an LLM to rewrite C code for NIST 800-22 randomness tests, a QuickSort algorithm and AES-128 into HLS-synthesizable c. The LLM iteratively transforms the C code guided by user prompts, implementing functions like streaming data and hardware-specific signals. This evaluation demonstrates the LLM's potential to assist hardware design  refactoring regular C code into HLS synthesizable C code. 
\end{abstract}
\noindent

\section{Introduction}
High-Level Synthesis (HLS) is a promising chip design methodology that enables rapid hardware design from high-level specifications.  HLS tools convert a high-level specification (C, C++) into an \ac{RTL} description~\cite{hls}: (1) HLS uses state-of-the-art compilers (e.g., LLVM or GCC) to extract a high-level control data flow graph (CDFG). (2) They then assign operations to time (scheduling) and space (allocation and binding) to determine the micro-architecture. HLS tools also support pragmas and directives to explore architectural choices  for a C specification.
Software and hardware paradigms are different. Hence HLS tools support a subset of C constructs that map to hardware. For instance, hardware does not support dynamic memory allocation and recursive constructs. Outputs can only communicate through parameters, arrays sizes need to be static, limit support for pointers and multiple processes can be modeled through independent function instances mapped into hardware blocks. Designers manually refactor C code to remove these constructs and make it compatible with HLS tools. However, manual refactoring is time consuming, and error prone~\cite{deeprajQPC}.

\subsection{Large Language Models (LLMs)} 
LLMs are trained on massive amounts of text data and excel at tasks like code generation and translation, particularly in languages like C, Cc++, and Python. However, their performance suffers on Hardware Description Languages (HDLs) like Verilog or VHDL due to the limited amount of training data available in those languages~\cite{hammond2022}.
This paper explores LLMs for refactoring C code, transforming it into a subset C compatible with HLS tools.  
LLMs can analyze generic C code and refactor it to HLS-synthesizable C. 
This  leverages the power of LLMs in generating and manipulating C code and 
bypassing their limitations in generating HDL code.
\subsection{Contributions and Roadmap}
\begin{figure}[!tb]
  \centering
  \includegraphics[width=\columnwidth]{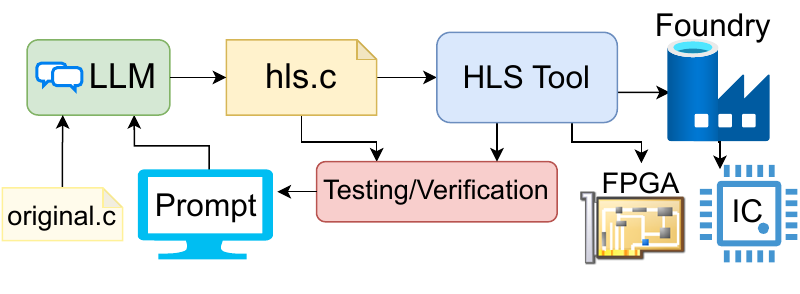}
  \vspace*{-15pt}
  \caption{Flow for the proposed C2HLSC approach.}
  \vspace*{-15pt}
  \label{fig:hlsflow}  
  
\end{figure}

LLMs have limited success in generating Verilog code~\cite{verigen,chipchat}. This is due to the scarcity online of hardware description language (HDL) data compared to code in C, C++ and other software. We explore use of LLMs to assist developers in refactoring generic C specification into synthesizable C that is compatible with HLS as shown in Figure~\ref{fig:hlsflow}. 
Out work has two main contributions. First, we present case studies of deriving HLS-usable C codes derived from generic C codes. The study compares development time and resource use achieved using LLM-assisted development in lieu of cumbersome manual methods. Second, we discuss the results and present a prototype hands-free C2HLSC LLM-based tool, which will be open-source. The paper road-map is as follows: 
\begin{enumerate*}
    \item Section \ref{sec:rel} presents related work, highlighting the novelties of the  approach.
    \item Section \ref{sec:cs} illustrates the case study, presenting the tasks and methodologies and results.
    \item Section \ref{sec:automated} presents the prototype hands-free C2HLSC tool, discussing capabilities and limitations.
    \item Section \ref{sec:conc} concludes the paper. 
\end{enumerate*}

\section{Related Work} \label{sec:rel}
Previous work  explored LLMs to design hardware. Verigen fine-tuned an LLM to improve its ability to produce Verilog~\cite{verigen}. The fine tuned LLM though  performed marginally better than ChatGPT3.5-turbo with an accuracy $\sim$65\%.  ChipChat~\cite{chipchat} was the first to tapeout a design written by an AI model. However, the single shot performance of the AI model was low and needed several iterations in order for the LLM to get to the correct result.
We target generating synthesizable C code as LLMs are more capable at C than at hardware languages~\cite{hammond2022}. In~\cite{amaranth} an LLM was used to write Amaranth HDL, a Python based HDL, that allows to model synchronous logic at the RTL. For this reason, while it uses a high level language, its semantics are close to verilog, and targets hardware designers. While the LLM came up with parts of the design, it fell short in some tasks, like generating interfaces. Software developers use HLS to design hardware and as such the code only provides the functionality\footnote{Whereas the hardware architecture and interface specification are instructed using HLS pragmas and directives.}.

\section{Case Study}\label{sec:cs}
\subsection{Overview}

We evaluated Gemini LLM~\cite{gemini} to transform C code into synthesizable C suitable for HLS. The evaluation consisted of two tasks. The {\bf first task} involved rewriting reference C implementations of the Frequency test, Frequency Block test, Cumulative Sums, and Overlapping Template Matching tests from the NIST 800-22 suite~\cite{nist} into synthesizable C code. These tests are designed to assess the randomness of a sequence. A {\bf first challenge} arose due to the inherent differences between software and hardware implementations. The reference C implementations operate on a pre-loaded random sequence stored in memory. Conversely, hardware implementations require on-the-fly analysis, processing the sequence bit-by-bit. This necessitates modifying the code to handle a streaming data input rather than a pre-loaded array. A {\bf second  challenge} stemmed from the p-value calculation. In software context, the precise p-value is critical and computed on-the-fly. However, since the hardware implementations primarily focus on distinguishing random from non-random sequences one can simplify this by pre-computing certain values offline and reducing the computational burden during on-the-fly analysis. Both these challenges
are non-trivial for  human developers and LLMs. 
The {\bf second task} assesses the LLM's ability to rewrite code constructs that are not supported by HLS tools. We used two algorithms: a QuickSort containing pointers and recursion~\cite{quicksort}, and the AES128 encrypt from the tinyAES library~\cite{tinyAES} with six functions. The goal was for the LLM to generate code without pointers and recursion, making it suitable for (Catapult) HLS.

\subsection{Methodology}
We broke down the process into small steps to allow the LLM to transform the original C into synthesizable C. For the first task we followed the following steps for the three tests: 
\begin{enumerate*}
    \item Present task to the LLM: "Hi, I have this code in C that I need to rewrite such that I can use it with an HLS tool to generate hardware.".
    \item Ask to remove print statements.
    \item Ask to rewrite the function as a streaming interface: "Now I need to rewrite the function such that it will get inferred as a streaming interface, to do so, I need to get rid of the epsilon array and have the function take a parameter to accept a bit at each function call."
    \item Ask to remove math steps to be computed offline (in some cases, ask to write a script to run them).
    \item Ask to add \textit{is\_random} and \textit{valid} signals as parameters.
    \item Ask to optimize data types using arbitrary width integers and fixed point arithmetic using HLSLIBS~\cite{hlslibs}.
    \item Ask to write a main function to test the  function passing random bits. 
    \item Ask to fix mistakes passing errors from  HLS tool.
\end{enumerate*}

For QuickSort we followed these steps:
\begin{enumerate*}
    \item Present the task to the LLM: "Hi, I have this code in C that I need to rewrite such that I can use it with an HLS tool to generate hardware.".
    \item Ask to remove print statements.
    \item Ask to rewrite function without using pointers.
    \item Ask to rewrite function without recursion.
    \item Ask to fix array sizes in function parameters.
    \item Ask to optimize data types using arbitrary width integers and fixed point arithmetic using HLSLIBS.
    \item Ask to write a main function to test the function passing an array to sort.
    \item Ask to fix mistakes by passing errors from HLS tool.
\end{enumerate*}

For the AES 128 from tinyAES~\cite{tinyAES} we followed the following steps asking to fix one function at a time:
\begin{enumerate*}
    \item Present the task to the LLM: "Hi, I have this code in C that I need to rewrite such that I can use it with an HLS tool to generate hardware.".
    \item Ask to rewrite for loops with fixed bounds and no pointer usage.
    \item Ask to rewrite the function parameters to using fixed size arrays.
    \item Ask to fix eventual mistakes passing errors from the HLS tool.
\end{enumerate*}
When the LLM responds with sub optimal answers, we check alternative answers, and if none fully satisfied the request we instruct the LLM with additional prompts including more details pointing out where the problem was, and, if not sufficient, hinting at possible solutions.

\subsection{Results} \label{sec:res}
The aim of this study is to evaluate how LLMs perform at rewriting C code such that it is HLS synthesizable\footnote{Finding pragmas to optimize the hardware architecture is an orthogonal problem for which LLMs could be employed, left as future work.}. We run the code through Catapult HLS to check correctness after synthesis, but we do not focus on the resource utilization, as it is depends on the architectural decisions. We targeted the nangate45 library at 50 MHz with a synchronous active high reset for all the tests.
The LLM was able to rewrite all C codes to run on Catapult HLS. We performed simulations with Modelsim to check result equivalency between the original C and the synthesized Verilog obtained from the LLM-generated C. The original C codes (without printfs) and the HLS C code rewritten by the LLM are  in Appendix~\ref{app:code}, links to the conversations are in Appendix~\ref{app:conv}. We can classify the errors in the LLM generated code into compile/synthesis errors, and functional errors. The former where easier to fix instructing the LLM with the error message and affected line. For functional errors caught by the failing tests, it was harder for the LLM to identify the cause and propose a fix. Manual effort was needed to point out sources of the issues for the LLM to propose an effective fix. Typical functional issues where mistaken loop readjustments, widths and  arbitrary widths types.

\subsubsection{C2HLSC of NIST 800-22 Randomness Tests}
The LLM performed well in most steps, especially in rewriting the function as a streaming interface, which is one of the more complex steps. Figure \ref{fig:streaming} shows the streaming interface rewriting for the Frequency Block test. 
For all the three tests this step was performed correctly at the first shot. However, the LLM did stumble on tasks that seemd more trivial. For example, when asked to remove the \texttt{abs} function in this line of code \texttt{(abs(sum.to\_int())  \textless= SUM\_THRESHOLD);} it came up with convoluted ways that did not work like \texttt{(sum.to\_int() \textless 0) ? -sum.to\_int() : sum.to\_int()) \textless= SUM\_THRESHOLD} and it took many tries to eventually get the expected result \texttt{(sum \textless SUM\_THRESHOLD) \&\& (sum \textless -SUM\_THRESHOLD)}. 
The test function generation was always generated effectively. 
\begin{figure}[tb]
  \centering
  \begin{subfigure}[t]{.9\columnwidth}
    \begin{lstlisting}
I need to rewrite the function such that it 
will get inferred as a streaming interface, 
to do so, I need to get rid of the epsilon 
array and have the function take a parameter
to accept a bit at each function call.    
\end{lstlisting}
\vspace{-7pt}
  \end{subfigure}
  \begin{subfigure}[t]{.9\columnwidth}
    \begin{lstlisting}[language=c]
void BlockFrequency(int M, int n) {
int i, j, N, blockSum;
double p_value, sum, pi, v, chi2;
N = n / M;  /* # OF SUBSTRING BLOCKS */
sum = 0.0;
for (i = 0; i < N; i++) {
    blockSum = 0;
    for (j = 0; j < M; j++) {
        blockSum += epsilon[j + i * M];
    }
    pi = (double)blockSum / (double)M;
    v = pi - 0.5;
    sum += v * v;
}
chi2 = 4.0 * M * sum;
p_value = cephes_igamc(N / 2.0, chi2 / 2.0);}
\end{lstlisting}
\vspace{-7pt}
  \end{subfigure}
  \begin{subfigure}[t]{.9\columnwidth}
    \begin{lstlisting}
Here's the code rewritten to accept a single
bit per function call, making it more suitable
for a streaming interface:
    \end{lstlisting}
    \vspace{-7pt}

  \end{subfigure}
  \begin{subfigure}[t]{.9\columnwidth}
    \begin{lstlisting}[language=c]
void BlockFrequency(int M, int n, bool bit) {
static int i = 0, j = 0, blockSum = 0;
static double sum = 0.0;
double pi, v, chi2;
if (j == 0) { // Reset on new block
    i++;
    blockSum = 0;
}
blockSum += bit; // Accumulate bit
j++;
if (j == M) { // Block complete
    pi = (double)blockSum / (double)M;
    v = pi - 0.5;
    sum += v * v;
    j = 0;
}
if (i == (n / M)) { // All blocks processed
    chi2 = 4.0 * M * sum;
    p_value=cephes_igamc(n / 2.0, chi2 / 2.0);
}}
\end{lstlisting}
\vspace{-7pt}
  \end{subfigure}
  \caption{Transforming into a streaming pipeline.}
  \label{fig:streaming}
\end{figure}

\subsubsection{C2HLSC of QuickSort}
The original QuickSort used a function to swap elements in the arrays using pointers. Inlining the function and removing pointers was flawless as reported in Figure \ref{fig:pointers}. However, in the subsequent steps it reverted to using the function and needed re-prompting to inline.
The LLM  got rid of recursion in one shot, using a fixed size stack to track the high and low values. However, the stack management logic was not correct. The problem was identified at the end, when running the generated test, which caused a segmentation fault. The LLM fixed the problem in a few tries by feeding the errors from gdb and showing intermediate values of high and low.

\begin{figure}[tb]
  \centering
  \begin{subfigure}[t]{.9\columnwidth}
    \begin{lstlisting}
Can we get rid of the swap function and replace
it inline? 
\end{lstlisting}
\vspace{-7pt}
  \end{subfigure}
  \begin{subfigure}[t]{.9\columnwidth}
    \begin{lstlisting}[language=c]
void swap(int* a, int* b) {
    int temp = *a;
    *a = *b;
    *b = temp;}
...
    if (i < j) {
        swap(arr[i], arr[j]); // Call to modified swap function
    }}
swap(&arr[low], &arr[j]);
...
\end{lstlisting}
\vspace{-7pt}
  \end{subfigure}
  \begin{subfigure}[t]{.9\columnwidth}
    \begin{lstlisting}
Yes, you can definitely get rid of the swap
function and replace it inline within the
partition function. Here's the modified code:
    \end{lstlisting}
    \vspace{-7pt}

  \end{subfigure}
  \begin{subfigure}[t]{.9\columnwidth}
    \begin{lstlisting}[language=c]
...
    if (i < j) {
        // Inline swap using temporary variable
        int temp = arr[i];
        arr[i] = arr[j];
        arr[j] = temp;
    }}
// Swap pivot element with element at index j
int temp = arr[low];
arr[low] = arr[j];
arr[j] = temp;
return j;
...
\end{lstlisting}
\vspace{-7pt}
  \end{subfigure}
  \caption{Removing pointers and inlining.}
  \label{fig:pointers}
\end{figure}

\subsubsection{C2HLSC of AES 128 Encrypt}
We fed one function at a time to the LLM starting from the top function \texttt{Cipher}. We tested the generated C code and found errors. We prompted the LLM to correct the errors. The LLM made major changes in the MixColumn and ShiftRows functions, so we asked it to double check that the new code was correct. ShiftRow function was corrected in 2 iterations, without extra context on the error. Results are  in Figure \ref{fig:aes}. Fixing MixColumn function took more iterations even though the error was simple.
\begin{figure}[tb]
  \centering
  \begin{subfigure}[t]{.9\columnwidth}
    \begin{lstlisting}
Can you fix possible problems in this function?
\end{lstlisting}
\vspace{-7pt}
  \end{subfigure}
  \begin{subfigure}[t]{.9\columnwidth}
    \begin{lstlisting}[language=c]
static void ShiftRows(state_t* state){
  uint8_t temp;
  // Rotate first row 1 columns to left  
  temp           = (*state)[0][1];
  (*state)[0][1] = (*state)[1][1];
  (*state)[1][1] = (*state)[2][1];
  (*state)[2][1] = (*state)[3][1];
  (*state)[3][1] = temp;
  // Rotate second row 2 columns to left  
  temp           = (*state)[0][2];
  (*state)[0][2] = (*state)[2][2];
  (*state)[2][2] = temp;
  temp           = (*state)[1][2];
  (*state)[1][2] = (*state)[3][2];
  (*state)[3][2] = temp;
  // Rotate third row 3 columns to left
  temp           = (*state)[0][3];
  (*state)[0][3] = (*state)[3][3];
  (*state)[3][3] = (*state)[2][3];
  (*state)[2][3] = (*state)[1][3];
  (*state)[1][3] = temp;}
...
\end{lstlisting}
\vspace{-7pt}
  \end{subfigure}
  \begin{subfigure}[t]{.9\columnwidth}
    \begin{lstlisting}
LLM result after 2 tries telling the LLM that 
the loop logic was wrong.
    \end{lstlisting}
    \vspace{-7pt}

  \end{subfigure}
  \begin{subfigure}[t]{.9\columnwidth}
    \begin{lstlisting}[language=c]
static void ShiftRows(state_t state[4][4]) {
uint8_t temp;
int i, j;
// Loop over rows 1 to 3 (0-based indexing)
for (i = 1; i < 4; ++i) {
// Number of positions to left shift for current row
  int shift = i;
  // Use a separate loop for the number of shifts
  for (j = 0; j < shift; ++j) {
    temp = state[0][i];  // Store the element to be shifted
    // Shift all elements by one position to the left
    for (int k = 0; k < 3; ++k) {
      state[k][i] = state[k + 1][i];
    }
  state[3][i] = temp; // Place the temp value at the end
}}}
\end{lstlisting}
\vspace{-7pt}
  \end{subfigure}
  \caption{Removing pointers and inlining.}
  \label{fig:aes}
\end{figure}

\subsection{Discussion}
Using LLMs for refactoring C code into HLS-compatible formats is a promising avenue in LLM aided design. The LLM, in our case Google's Gemini, refactored reference C implementations of NIST 800-22 randomness tests, QuickSort, and AES-128. The LLM's ability to iteratively refactor the code based on user instructions significantly reduces manual effort in adapting code for HLS tools. This can streamline the hardware design process, especially for repetitive tasks. The LLM effectively addressed challenges like converting code from memory-based data processing to streaming, from recursion to iteration and pointers. While the LLM achieved core functionalities, it occasionally struggled with minor details requiring several iterations to guide it to the correct solution. In a practical scenario, a developer can rectify these minor errors. However, for an automated flow, a feedback loop is crucial, like that in~\cite{chipchat}.

\begin{table}[!bht]
\centering
\caption{Resource Utilization and Latency Results}
\label{tab:area}
\resizebox{\columnwidth}{!}{%
\begin{tabular}{@{}lrrrrrr@{}}
\toprule
\multirow{2}{*}{Design} & \multicolumn{2}{c}{Area Score}                             & \multicolumn{2}{c}{\# Operations}                                  & \multicolumn{2}{c}{Latency}                                \\
                        & \multicolumn{1}{l}{This work} & \multicolumn{1}{l}{Manual} & \multicolumn{1}{l}{This work} & \multicolumn{1}{l}{Manual} & \multicolumn{1}{l}{This work} & \multicolumn{1}{l}{Manual} \\ \midrule
NIST-Monobit  & 244  & 225.3  & 19 & 19  & 1 & 1  \\
NIST-Monobit Block  & 702.3   & 826.0  & 24  & 20  & 1  & 1  \\
NIST-Cusums  & 677.4  & 632  & 24 & 28 & 1   & 1  \\
NIST-Overlapping & 9933.4 & 7172.1  & 165 &  118 & 1 & 1 \\ 
QuickSort  & 18115.8 & n.a.  & 67 & n.a. & 18   & n.a. \\
AES & 38604.5  & n.a.  & 1924 & n.a. & 160  & n.a. \\ \bottomrule
\end{tabular}%
}
\end{table}

Table \ref{tab:area} shows area for the implemented designs. For NIST test implementation we have reference designs that were implemented by a graduate student. We used the same directives for a fair comparison between the 2. Area scores from Catapult are close. The manual implementations took around 4 hours each while C2HLSC took between 30 to 60 minutes each. Although the sample size is limited, this shows the potential of LLMs in speeding up the process effectively and efficiently.

\section{Hands-Free C2HSLC Prototype}\label{sec:automated}
\begin{figure}[!htb]
  \centering
  \includegraphics[width=\columnwidth]{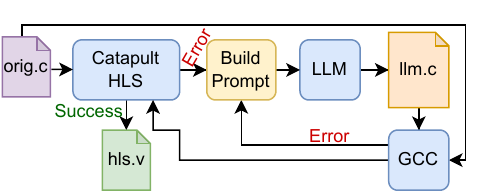}
  \caption{Hands-free C2HSLC LLM-based prototype flow.}
  \label{fig:prototype}
\end{figure}

With the experience of the engineer-in-the-loop case studies, we implemented a hands-free C2HLSC prototype. The first insight from the case study was the twofold nature of the errors that can occur in the generated C --- functional/compile errors and synthesis errors. The former kind can be caught by gcc. The latter kind can be caught by running the Catapult HLS tool. For this reason we setup a double feedback loop as shown in Figure~\ref{fig:prototype}. One checks that the generated code compiles and passes reference tests, and one that checks that the code is synthesizable by Catapult HLS. We selected ChatGPT 3.5 Turbo as we did not have access to Gemini APIs. The flow is implemented in Python and will be made open source.

We ran the flow on the QuickSort and AES C codes used in the case study. In the current version, the flow can handle only a limited number of functions at a time and so for AES we broke the code down into the individual functions.

\begin{table}[htb]
\centering
\caption{Hands-free C2HLSC LLM-based flow.}
\label{tab:flow}
\resizebox{\columnwidth}{!}{%
\begin{tabular}{@{}llrrr@{}}
\toprule
Design & \# Prompts & \multicolumn{1}{c}{Area Score} & \multicolumn{1}{c}{\# Ops} & \multicolumn{1}{c}{Latency} \\ \midrule
QuickSort   & 6 & 52184 & 1961 & 8   \\
AddRoundKey & 9 & 702.3 & 837  & 210 \\
ShiftRows   & 1 & 23.9  & 3    & 0   \\
MixColumns  & 1 &  635.9  & 225  & 6   \\
SubBytes    & 1 &  1176.5  &   53   &   33  \\ \bottomrule
\end{tabular}%
}
\end{table}

The flow was able to automatically rewrite the functions without any human intervention. We plan to expand it by allowing more functions at a time, and include better optimizations. The generated codes are in Appendix \ref{app:code}.

\section{Conclusions} \label{sec:conc}
The case studies demonstrated the potential of LLMs in aiding HLS. 
The implemented prototype shows the potential of LLMs in aiding hardware design leveraging HLS.
Future work includes the use of LLMs to discover architectural trade-offs by applying pragmas and directives. Together with the proposed flow, it will enable a seamless transition from C to HLS C to optimized hardware architectures.

\newpage
\bibliographystyle{IEEEtran}
\bibliography{main.bib}

\newpage
\appendix

\subsection{LLM generated HLS compatible C codes} \label{app:code}
In this section we report original codes and final codes of our case study. We cut print statements from the originals and comments from the results.

\begin{figure}[tb]
  \centering
  \begin{subfigure}[t]{.9\columnwidth}
    \begin{lstlisting}
Original c code (after removing printf s)   \end{lstlisting}
\vspace{-7pt}
  \end{subfigure}
  \begin{subfigure}[t]{.9\columnwidth}
    \begin{lstlisting}[language=c]
void Frequency(int n)
{
	int		i;
	double	f, s_obs, p_value, sum, sqrt2 = 1.41421356237309504880;
	sum = 0.0;
	for ( i=0; i<n; i++ )
		sum += 2*(int)epsilon[i]-1;
	s_obs = fabs(sum)/sqrt(n);
	f = s_obs/sqrt2;
	p_value = erfc(f);
}
\end{lstlisting}
\vspace{-7pt}
  \end{subfigure}
  \begin{subfigure}[t]{.9\columnwidth}
    \begin{lstlisting}
HLS compatible code rewritten by the LLM
    \end{lstlisting}
    \vspace{-7pt}

  \end{subfigure}
  \begin{subfigure}[t]{.9\columnwidth}
    \begin{lstlisting}[language=c]
#define TARGET_BITS 128
#define P_VALUE_THRESHOLD 0.01
#define THRESHOLD 29  
typedef ac_int<9, true> sum_type;  // true for signed
typedef ac_int<8, false> count_type;  // false for unsigned
typedef ac_int<1, false> bit_type;
void
Frequency_Bit(bit_type *bit, bit_type *valid, bit_type *result)
{
  static sum_type    sum = 0;
  static count_type count = 0;
  sum += (*bit) ? sum_type(1) : sum_type(-1); 
  count++;
  if (count == TARGET_BITS) {
    // Check if sum is within the threshold range
    *result = (sum <= THRESHOLD) && (sum >= -THRESHOLD); 
    *valid = 1;  
    count = 0;
    sum = 0;
  } else {
    *result = 0;
    *valid = 0;
}}
\end{lstlisting}
\vspace{-7pt}
  \end{subfigure}
  \caption{Human-in-the-loop Monobit}
  \label{fig:monobit}
\end{figure}

\begin{figure}[tb]
  \centering
  \begin{subfigure}[t]{.9\columnwidth}
    \begin{lstlisting}
Original c code (after removing printf s)   \end{lstlisting}
\vspace{-7pt}
  \end{subfigure}
  \begin{subfigure}[t]{.9\columnwidth}
    \begin{lstlisting}[language=c]
void BlockFrequency(int M, int n)
{
	int		i, j, N, blockSum;
	double	p_value, sum, pi, v, chi_squared;
	N = n/M; 		/* # OF SUBSTRING BLOCKS      */
	sum = 0.0;
	
	for ( i=0; i<N; i++ ) {
		blockSum = 0;
		for ( j=0; j<M; j++ )
			blockSum += epsilon[j+i*M];
		pi = (double)blockSum/(double)M;
		v = pi - 0.5;
		sum += v*v;
	}
	chi_squared = 4.0 * M * sum;
	p_value = cephes_igamc(N/2.0, chi_squared/2.0);
}
\end{lstlisting}
\vspace{-7pt}
  \end{subfigure}
  \begin{subfigure}[t]{.9\columnwidth}
    \begin{lstlisting}
HLS compatible code rewritten by the LLM
    \end{lstlisting}
    \vspace{-7pt}

  \end{subfigure}
  \begin{subfigure}[t]{.9\columnwidth}
    \begin{lstlisting}[language=c]
void BlockFrequency(ac_int<1, false> bit, ac_int<1, false>& valid, ac_int<1, false>& result) {
  static ac_int<7, false> blockSum = 0;
  static ac_fixed<16, 8, false> sum = 0.0; 
  ac_fixed<8, 4, false> pi_fixed, v;
  static ac_int<4, false> i = 0;
  static ac_int<8, false> j = 0;
  if (j == 0) {
    i++;
    blockSum = 0;
  }
  blockSum += bit; 
  j++;
  if (j == BLOCK_SIZE) { 
    pi_fixed = (ac_fixed<12, 4, false>) blockSum / BLOCK_SIZE;
    const ac_fixed<8, 4, false> half = 0.5;  
    v = pi_fixed - half;
    sum += v * v;
    j = 0;
  }
  if (i == NUM_BLOCKS) { 
    const ac_fixed<16, 8, false> threshold = 13.8155 / 512.0;  
    valid = true;  
    result = (sum < threshold) ? 0 : 1;
  } else {
    valid = false; 
    result = 0;  
}}
\end{lstlisting}
\vspace{-7pt}
  \end{subfigure}
  \caption{Human-in-the-loop Monobit Block}
  \label{fig:monobit}
\end{figure}

\begin{figure}[tb]
  \centering
  \begin{subfigure}[t]{.9\columnwidth}
    \begin{lstlisting}
Original c code (after removing printf s)   \end{lstlisting}
\vspace{-7pt}
  \end{subfigure}
  \begin{subfigure}[t]{.9\columnwidth}
    \begin{lstlisting}[language=c]
void CumulativeSums(int n) {
    int S, sup, inf, z, zrev, k;
    double sum1, sum2, p_value;
    S = 0; sup = 0; inf = 0;
    for (k = 0; k < n; k++) {
        epsilon[k] ? S++ : S--;
        if (S > sup) sup++;
        if (S < inf) inf--;
        z = (sup > -inf) ? sup : -inf;
        zrev = (sup - S > S - inf) ? sup - S : S - inf;}
    sum1 = 0.0; // forward
    for (k = (-n / z + 1) / 4; k <= (n / z - 1) / 4; k++) {
        sum1 += cephes_normal(((4 * k + 1) * z) / sqrt(n));
        sum1 -= cephes_normal(((4 * k - 1) * z) / sqrt(n));}
    sum2 = 0.0;
    for (k = (-n / z - 3) / 4; k <= (n / z - 1) / 4; k++) {
        sum2 += cephes_normal(((4 * k + 3) * z) / sqrt(n));
        sum2 -= cephes_normal(((4 * k + 1) * z) / sqrt(n));}
    p_value = 1.0 - sum1 + sum2;
    sum1 = 0.0; // backwards
    for (k = (-n / zrev + 1) / 4; k <= (n / zrev - 1) / 4; k++) {
        sum1 += cephes_normal(((4 * k + 1) * zrev) / sqrt(n));
        sum1 -= cephes_normal(((4 * k - 1) * zrev) / sqrt(n));}
    sum2 = 0.0;
    for (k = (-n / zrev - 3) / 4; k <= (n / zrev - 1) / 4; k++) {
        sum2 += cephes_normal(((4 * k + 3) * zrev) / sqrt(n));
        sum2 -= cephes_normal(((4 * k + 1) * zrev) / sqrt(n));}
    p_value = 1.0 - sum1 + sum2;}
\end{lstlisting}
\vspace{-7pt}
  \end{subfigure}
  \begin{subfigure}[t]{.9\columnwidth}
    \begin{lstlisting}
HLS compatible code rewritten by the LLM
    \end{lstlisting}
    \vspace{-7pt}

  \end{subfigure}
  \begin{subfigure}[t]{.9\columnwidth}
    \begin{lstlisting}[language=c]
#define N  20000  
#define THRESHOLD (ac_int<12, false>)397
void CumulativeSums_Streaming(ac_int<1, false> bit, ac_int<1, false>* output, ac_int<1, false>* valid) {
  static ac_int<12, false> S = 0; 
  static ac_int<12, false> sup = 0;
  static ac_int<12, false> inf = 0;
  S += bit ? 1 : -1;
  sup = sup > S ? sup : S;  
  inf = inf < S ? inf : S;  
  // Logic for output based on sup and inf
  static ac_int<16, false> processed_bits = 0;  
  processed_bits++;
  if (processed_bits == N) {
    *output = (sup < THRESHOLD && inf > -THRESHOLD) ? 1 : 0;
    *valid = (processed_bits == N);  
    processed_bits = 0; 
  } else {
    *output = 0;
    *valid = 0;
}}
\end{lstlisting}
\vspace{-7pt}
  \end{subfigure}
  \caption{Human-in-the-loop Cusums}
  \label{fig:cusums}
\end{figure}

\begin{figure}[tb]
  \centering
  \begin{subfigure}[t]{.9\columnwidth}
    \begin{lstlisting}
Original c code (after removing printf s)   \end{lstlisting}
\vspace{-7pt}
  \end{subfigure}
  \begin{subfigure}[t]{.9\columnwidth}
    \begin{lstlisting}[language=c]
void OverlappingTemplateMatchings(int m, int n)
{
int				i, k, match;
double			W_obs, eta, sum, chi2, p_value, lambda;
int				M, N, j, K = 5;
unsigned int	nu[6] = { 0, 0, 0, 0, 0, 0 };
//double			pi[6] = { 0.143783, 0.139430, 0.137319, 0.124314, 0.106209, 0.348945 };
double			pi[6] = { 0.364091, 0.185659, 0.139381, 0.100571, 0.0704323, 0.139865 };
BitSequence		*sequence;
M = 1032;
N = n/M;
if ( (sequence = (BitSequence *) calloc(m, sizeof(BitSequence))) == NULL ) {
    // ERROR
}
else
    for ( i=0; i<m; i++ )
        sequence[i] = 1;
lambda = (double)(M-m+1)/pow(2,m);
eta = lambda/2.0;
sum = 0.0;
for ( i=0; i<K; i++ ) {			/* Compute Probabilities */
    pi[i] = Pr(i, eta);
    sum += pi[i];
}
pi[K] = 1 - sum;

for ( i=0; i<N; i++ ) {
  W_obs = 0;
  for ( j=0; j<M-m+1; j++ ) {
    match = 1;
    for ( k=0; k<m; k++ ) {
     if ( sequence[k] != epsilon[i*M+j+k] )
        match = 0;
     }
     if ( match == 1 )
       W_obs++;
    }
    if ( W_obs <= 4 )
     nu[(int)W_obs]++;
    else
     nu[K]++;
}
sum = 0;
chi2 = 0.0;                                   /* Compute Chi Square */
for ( i=0; i<K+1; i++ ) {
    chi2 += pow((double)nu[i] - (double)N*pi[i], 2)/((double)N*pi[i]);
    sum += nu[i];
}
p_value = cephes_igamc(K/2.0, chi2/2.0);
}
\end{lstlisting}
\vspace{-7pt}
  \end{subfigure}

    \caption{Overlapping Original}
  \label{fig:monobit}
\end{figure}

\begin{figure}[tb]
  \centering
    \begin{subfigure}[t]{.9\columnwidth}
    \begin{lstlisting}
HLS compatible code rewritten by the LLM
    \end{lstlisting}
    \vspace{-7pt}

  \end{subfigure}
  \begin{subfigure}[t]{.9\columnwidth}
    \begin{lstlisting}[language=c]
#define N 1048576  
#define M 1032  
#define N_WIDTH 20  
#define M_WIDTH 10  
#define CHI2_THRESHOLD_FIXED ac_fixed<21, 21, false>(1056300.0)  
void OverlappingTemplateMatchings(int epsilon, ac_int<1, false> *is_random_out, ac_int<1, false> *valid_output_out) {
  static ac_int<9, false> W_obs = 0;  
  static ac_int<10, false> nu[6] = {0, 0, 0, 0, 0, 0};  
  static ac_int<9, false> shift_reg[9] = {0}; 
  static ac_int<N_WIDTH+1, false> n_counter = 0; 
  static const ac_fixed<32, 16, false> pi[6] = {
    2.746566, 
    ac_fixed<32, 16, false>(5.386218), ac_fixed<32, 16, false>(7.17457),
    ac_fixed<32, 16, false>(9.94322), ac_fixed<32, 16, false>(14.198031), ac_fixed<32, 16, false>(7.1497515)
  };
ac_fixed<23, 22> chi2 = 0;
  for (int i = 8; i > 0; i--) {
    shift_reg[i] = shift_reg[i - 1];
  }
  shift_reg[0] = ac_int<1, false>(epsilon);  
  ac_int<1, false> match = 1;
  for (int i = 0; i < 9; i++) {
    if (shift_reg[i] != 1) { 
      match = 0;
      break;
    }
  }
  if (match) {
    W_obs++;
  }
  n_counter++;
  if (n_counter % M == M-1) {
    if (W_obs <= 4) {
      nu[(int)W_obs]++;
    } else {
      nu[5]++;  
    }
    W_obs = 0; 
  }
  *is_random_out = 0;
  *valid_output_out = 0;
  if (n_counter == N) {
    for (int i = 0; i < 6; i++) {
        chi2 +=  nu[i] * nu [i] * pi[i];
    }
    *is_random_out = chi2 < CHI2_THRESHOLD_FIXED; 
  *valid_output_out = true;
    for (int i = 0; i < 6; i++) {
      nu[i] = 0;
    }
    n_counter = 0;  
}}
\end{lstlisting}
\vspace{-7pt}
  \end{subfigure}
  \caption{Human-in-the-loop Overlapping}
  \label{fig:monobit}
\end{figure}

\begin{figure}[tb]
  \centering
  \begin{subfigure}[t]{.9\columnwidth}
    \begin{lstlisting}
Original c code (after removing printf s)   \end{lstlisting}
\vspace{-7pt}
  \end{subfigure}
  \begin{subfigure}[t]{.9\columnwidth}
    \begin{lstlisting}[language=c]
void swap(int* a, int* b) { 
    int temp = *a; 
    *a = *b; 
    *b = temp; } 
  int partition(int arr[], int low, int high) { 
    int pivot = arr[low]; 
    int i = low; 
    int j = high; 
    while (i < j) { 
        while (arr[i] <= pivot && i <= high - 1) { 
            i++; 
        } 
        while (arr[j] > pivot && j >= low + 1) { 
            j--; 
        } 
        if (i < j) { 
            swap(&arr[i], &arr[j]); 
        } 
    } 
    swap(&arr[low], &arr[j]); 
    return j; } 
void quickSort(int arr[], int low, int high) { 
    if (low < high) { 
        int partitionIndex = partition(arr, low, high); 
        quickSort(arr, low, partitionIndex - 1); 
        quickSort(arr, partitionIndex + 1, high); 
} } 
\end{lstlisting}
\vspace{-7pt}
  \end{subfigure}

    \caption{Quicksort original}
  \label{fig:quick}
\end{figure}

\begin{figure}[tb]
  \centering
  \begin{subfigure}[t]{.9\columnwidth}
    \begin{lstlisting}
HLS compatible code rewritten by the LLM
    \end{lstlisting}
    \vspace{-7pt}

  \end{subfigure}
  \begin{subfigure}[t]{.9\columnwidth}
    \begin{lstlisting}[language=c]
int partition(int arr[MAX_SIZE], int low, int high) {}
    int pivot = arr[low];
    int i = low;
    int j = high;
    if (low == high) {
        return low;
    }
    while (i < j) {
        while (arr[i] <= pivot && i <= high - 1) {
            i++;
        }
        while (arr[j] > pivot && j >= low + 1) {
            j--;
        }
        if (i < j) {
            int temp = arr[i];
            arr[i] = arr[j];
            arr[j] = temp;
        }
    }
    int temp = arr[low];
    arr[low] = arr[j];
    arr[j] = temp;
    return j;
}
void quickSort(int arr[MAX_SIZE], int low, int high) {
int stack[100]; 
int top = -1;
if (high - low + 1 > MAX_SIZE) {
    return;
}
stack[++top] = low;
stack[++top] = high;
while (top >= 0) {
    high = stack[top--];
    low = stack[top--];
    int partitionIndex = partition(arr, low, high);
    if (partitionIndex + 1 < high) {
        stack[++top] = partitionIndex + 1;
        stack[++top] = high;
    }
    if (low < partitionIndex - 1) {
        stack[++top] = low;
        stack[++top] = partitionIndex - 1;
}}}
\end{lstlisting}
\vspace{-7pt}
  \end{subfigure}
  \caption{Human-in-the-loop Quicksort}
  \label{fig:quick2}
\end{figure}

\begin{figure}[tb]
  \centering
  \begin{subfigure}[t]{.9\columnwidth}
    \begin{lstlisting}
Original c code (after removing printf s)   \end{lstlisting}
\vspace{-7pt}
  \end{subfigure}
  \begin{subfigure}[t]{.9\columnwidth}
    \begin{lstlisting}[language=c]
static void Cipher(state_t* state, const uint8_t* RoundKey)
{
  uint8_t round = 0;
  AddRoundKey(0, state, RoundKey);
  for (round = 1; ; ++round)
  {
    SubBytes(state);
    ShiftRows(state);
    if (round == Nr) {
      break;
    }
    MixColumns(state);
    AddRoundKey(round, state, RoundKey);
  }
  AddRoundKey(Nr, state, RoundKey);
}
\end{lstlisting}
\vspace{-7pt}
  \end{subfigure}
  \begin{subfigure}[t]{.9\columnwidth}
    \begin{lstlisting}
HLS compatible code rewritten by the LLM
    \end{lstlisting}
    \vspace{-7pt}

  \end{subfigure}
  \begin{subfigure}[t]{.9\columnwidth}
    \begin{lstlisting}[language=c]
void Cipher(state_t state[4][4], const uint8_t RoundKey[AES_keyExpSize]) {
  uint8_t round;
  AddRoundKey(0, state, RoundKey);
  for (round = 1; round <= Nr; ++round) {
    SubBytes(state);
    ShiftRows(state);
    if (round < Nr) { 
      MixColumns(state);
    }
    AddRoundKey(round, state, RoundKey);
  }
}
\end{lstlisting}
\vspace{-7pt}
  \end{subfigure}
  \caption{Human-in-the-loop AES Cipher}
  \label{fig:monobit}
\end{figure}

\begin{figure}[tb]
  \centering
  \begin{subfigure}[t]{.9\columnwidth}
    \begin{lstlisting}
Original c code (after removing printf s)   \end{lstlisting}
\vspace{-7pt}
  \end{subfigure}
  \begin{subfigure}[t]{.9\columnwidth}
    \begin{lstlisting}[language=c]
static void MixColumns(state_t* state)
{
  uint8_t i;
  uint8_t Tmp, Tm, t;
  for (i = 0; i < 4; ++i)
  {  
    t   = (*state)[i][0];
    Tmp = (*state)[i][0] ^ (*state)[i][1] ^ (*state)[i][2] ^ (*state)[i][3] ;
    Tm  = (*state)[i][0] ^ (*state)[i][1] ; 
    Tm = xtime(Tm);  (*state)[i][0] ^= Tm ^ Tmp ;
    Tm  = (*state)[i][1] ^ (*state)[i][2] ; 
    Tm = xtime(Tm);  (*state)[i][1] ^= Tm ^ Tmp ;
    Tm  = (*state)[i][2] ^ (*state)[i][3] ; 
    Tm = xtime(Tm);  (*state)[i][2] ^= Tm ^ Tmp ;
    Tm  = (*state)[i][3] ^ t ;              
    Tm = xtime(Tm);  (*state)[i][3] ^= Tm ^ Tmp ;
  }
}

\end{lstlisting}
\vspace{-7pt}
  \end{subfigure}
  \begin{subfigure}[t]{.9\columnwidth}
    \begin{lstlisting}
HLS compatible code rewritten by the LLM
    \end{lstlisting}
    \vspace{-7pt}

  \end{subfigure}
  \begin{subfigure}[t]{.9\columnwidth}
    \begin{lstlisting}[language=c]
static void MixColumns(state_t state[4][4]) {
  uint8_t i, Tmp, Tm, t;
  for (i = 0; i < 4; ++i) {
    t   = state[i][0];
    Tmp = state[i][0] ^ state[i][1] ^ state[i][2] ^ state[i][3];
    Tm  = state[i][0] ^ state[i][1]; Tm = xtime(Tm);
    state[i][0] ^= Tm ^ Tmp;
    
    // Corrected lines with direct array access
    Tm  = state[i][1] ^ state[i][2]; Tm = xtime(Tm);
    state[i][1] ^= Tm ^ Tmp;
    Tm  = state[i][2] ^ state[i][3]; Tm = xtime(Tm);
    state[i][2] ^= Tm ^ Tmp;
    Tm  = state[i][3] ^ t;              Tm = xtime(Tm);
    state[i][3] ^= Tm ^ Tmp;
  }
}
\end{lstlisting}
\vspace{-7pt}
  \end{subfigure}
  \caption{Human-in-the-loop AES Mix Columns}
  \label{fig:monobit}
\end{figure}

\begin{figure}[tb]
  \centering
  \begin{subfigure}[t]{.9\columnwidth}
    \begin{lstlisting}
Original c code (after removing printf s)   \end{lstlisting}
\vspace{-7pt}
  \end{subfigure}
  \begin{subfigure}[t]{.9\columnwidth}
    \begin{lstlisting}[language=c]
static void ShiftRows(state_t* state)
{
  uint8_t temp;

  // Rotate first row 1 columns to left  
  temp           = (*state)[0][1];
  (*state)[0][1] = (*state)[1][1];
  (*state)[1][1] = (*state)[2][1];
  (*state)[2][1] = (*state)[3][1];
  (*state)[3][1] = temp;

  // Rotate second row 2 columns to left  
  temp           = (*state)[0][2];
  (*state)[0][2] = (*state)[2][2];
  (*state)[2][2] = temp;

  temp           = (*state)[1][2];
  (*state)[1][2] = (*state)[3][2];
  (*state)[3][2] = temp;

  // Rotate third row 3 columns to left
  temp           = (*state)[0][3];
  (*state)[0][3] = (*state)[3][3];
  (*state)[3][3] = (*state)[2][3];
  (*state)[2][3] = (*state)[1][3];
  (*state)[1][3] = temp;
}
\end{lstlisting}
\vspace{-7pt}
  \end{subfigure}
  \begin{subfigure}[t]{.9\columnwidth}
    \begin{lstlisting}
HLS compatible code rewritten by the LLM
    \end{lstlisting}
    \vspace{-7pt}

  \end{subfigure}
  \begin{subfigure}[t]{.9\columnwidth}
    \begin{lstlisting}[language=c]
static void ShiftRows(state_t state[4][4]) {
  uint8_t temp;
  int i, j;

  // Loop over rows 1 to 3 (0-based indexing)
  for (i = 1; i < 4; ++i) {
    // Number of positions to left shift for current row
    int shift = i;

    // Use a separate loop for the number of shifts
    for (j = 0; j < shift; ++j) {
      temp = state[0][i];  // Store the element to be shifted
      // Shift all elements by one position to the left
      for (int k = 0; k < 3; ++k) {
        state[k][i] = state[k + 1][i];
      }
      state[3][i] = temp; // Place the temp value at the end
    }
  }
}
\end{lstlisting}
\vspace{-7pt}
  \end{subfigure}
  \caption{Human-in-the-loop AES Shift Rows}
  \label{fig:monobit}
\end{figure}

\begin{figure}[tb]
  \centering
  \begin{subfigure}[t]{.9\columnwidth}
    \begin{lstlisting}
Original c code (after removing printf s)   \end{lstlisting}
\vspace{-7pt}
  \end{subfigure}
  \begin{subfigure}[t]{.9\columnwidth}
    \begin{lstlisting}[language=c]
static void SubBytes(state_t* state)
{
  uint8_t i, j;
  for (i = 0; i < 4; ++i)
  {
    for (j = 0; j < 4; ++j)
    {
      (*state)[j][i] = getSBoxValue((*state)[j][i]);
    }
  }
}
\end{lstlisting}
\vspace{-7pt}
  \end{subfigure}
  \begin{subfigure}[t]{.9\columnwidth}
    \begin{lstlisting}
HLS compatible code rewritten by the LLM
    \end{lstlisting}
    \vspace{-7pt}

  \end{subfigure}
  \begin{subfigure}[t]{.9\columnwidth}
    \begin{lstlisting}[language=c]
static void SubBytes(state_t state[4][4]) {
  uint8_t i, j;
  for (i = 0; i < 4; ++i) {
    for (j = 0; j < 4; ++j) {
      state[j][i] = getSBoxValue(state[j][i]);
    }
  }
\end{lstlisting}
\vspace{-7pt}
  \end{subfigure}
  \caption{Human-in-the-loop AES Sub Bytes}
  \label{fig:monobit}
\end{figure}

\begin{figure}[tb]
  \centering
  \begin{subfigure}[t]{.9\columnwidth}
    \begin{lstlisting}
Original c code (after removing printf s)   \end{lstlisting}
\vspace{-7pt}
  \end{subfigure}
  \begin{subfigure}[t]{.9\columnwidth}
    \begin{lstlisting}[language=c]
static void AddRoundKey(uint8_t round, state_t* state, const uint8_t* RoundKey)
{
  uint8_t i,j;
  for (i = 0; i < 4; ++i)
  {
    for (j = 0; j < 4; ++j)
    {
      (*state)[i][j] ^= RoundKey[(round * Nb * 4) + (i * Nb) + j];
    }
  }
}
\end{lstlisting}
\vspace{-7pt}
  \end{subfigure}
  \begin{subfigure}[t]{.9\columnwidth}
    \begin{lstlisting}
HLS compatible code rewritten by the LLM
    \end{lstlisting}
    \vspace{-7pt}

  \end{subfigure}
  \begin{subfigure}[t]{.9\columnwidth}
    \begin{lstlisting}[language=c]
static void AddRoundKey(uint8_t round, state_t state[4][4], const uint8_t RoundKey[AES_KEYLEN]) {
  uint8_t i, j;
  uint8_t base = round * Nb * 4;
  for (i = 0; i < 4; ++i) {
    for (j = 0; j < 4; ++j) {
      state[i][j] ^= RoundKey[base + (i * Nb) + j];
    }
  }
}
\end{lstlisting}
\vspace{-7pt}
  \end{subfigure}
  \caption{Human-in-the-loop AES Add Round Key}
  \label{fig:monobit}
\end{figure}

\begin{figure}[tb]
  \centering
  \begin{subfigure}[t]{.9\columnwidth}
    \begin{lstlisting}
HLS compatible code from the hands-free flow
    \end{lstlisting}
    \vspace{-7pt}

  \end{subfigure}
  \begin{subfigure}[t]{.9\columnwidth}
    \begin{lstlisting}[language=c]
// Function to swap two elements
void swap(int* a, int* b){
int temp = *a;
*a = *b;
*b = temp;}
int partition(int arr[], int low, int high){
int pivot = arr[low];
int i = low;
int j = high;
while (i < j){
    while (arr[i] <= pivot && i <= high - 1){
        i++;
    }
    while (arr[j] > pivot && j >= low + 1){
        j--;
    }
    if (i < j){
        swap(&arr[i], &arr[j]);
    }
}
swap(&arr[low], &arr[j]);
return j;
}
void quickSort(int arr[64], int low, int high){
int stack[128]; // Fixed size stack to work with HLS tools
int top = -1;
stack[++top] = low;
stack[++top] = high;
while (top >= 0){
    high = stack[top--];
    low = stack[top--];
    if (low < high){
        int partitionIndex = partition(arr, low, high);
        stack[++top] = low;
        stack[++top] = partitionIndex - 1;
        stack[++top] = partitionIndex + 1;
        stack[++top] = high;
}}}
\end{lstlisting}
\vspace{-7pt}
  \end{subfigure}
  \caption{Hands-free C2HLSC QuickSort}
  \label{fig:monobit}
\end{figure}

\begin{figure}[tb]
  \centering
  \begin{subfigure}[t]{.9\columnwidth}
    \begin{lstlisting}
HLS compatible code from the hands-free flow
    \end{lstlisting}
    \vspace{-7pt}

  \end{subfigure}
  \begin{subfigure}[t]{.9\columnwidth}
    \begin{lstlisting}[language=c]
void AddRoundKey(uint8_t round, state_t state, const uint8_t* RoundKey)
{
uint8_t i, j;
uint8_t RoundKey_local[AES_keyExpSize];
for (i = 0; i < AES_keyExpSize; ++i)
{
#pragma HLS PIPELINE II=1
    RoundKey_local[i] = RoundKey[i];
}
for (i = 0; i < 4; ++i)
{
    for (j = 0; j < 4; ++j)
    {
#pragma HLS PIPELINE II=1
        state[i][j] ^= RoundKey_local[(round * Nb * 4) + (i * Nb) + j];
    }
}
}
\end{lstlisting}
\vspace{-7pt}
  \end{subfigure}
  \caption{Hands-free C2HLSC AES Add Round Key}
  \label{fig:monobit}
\end{figure}

\begin{figure}[tb]
  \centering
  \begin{subfigure}[t]{.9\columnwidth}
    \begin{lstlisting}
HLS compatible code from the hands-free flow
    \end{lstlisting}
    \vspace{-7pt}

  \end{subfigure}
  \begin{subfigure}[t]{.9\columnwidth}
    \begin{lstlisting}[language=c]
void ShiftRows(state_t state) {
    uint8_t temp;

    // Rotate first row 1 column to the left
    temp = state[0][1];
    state[0][1] = state[1][1];
    state[1][1] = state[2][1];
    state[2][1] = state[3][1];
    state[3][1] = temp;

    // Rotate second row 2 columns to the left
    temp = state[0][2];
    state[0][2] = state[2][2];
    state[2][2] = temp;

    temp = state[1][2];
    state[1][2] = state[3][2];
    state[3][2] = temp;

    // Rotate third row 3 columns to the left
    temp = state[0][3];
    state[0][3] = state[3][3];
    state[3][3] = state[2][3];
    state[2][3] = state[1][3];
    state[1][3] = temp;
}
\end{lstlisting}
\vspace{-7pt}
  \end{subfigure}
  \caption{Hands-free C2HLSC AES Shift Rows}
  \label{fig:monobit}
\end{figure}

\begin{figure}[tb]
  \centering
  \begin{subfigure}[t]{.9\columnwidth}
    \begin{lstlisting}
HLS compatible code from the hands-free flow
    \end{lstlisting}
    \vspace{-7pt}

  \end{subfigure}
  \begin{subfigure}[t]{.9\columnwidth}
    \begin{lstlisting}[language=c]
void MixColumns(state_t state)
{
    uint8_t i;
    uint8_t Tmp, Tm, t;
    for (i = 0; i < 4; ++i)
    {  
        t   = state[i][0];
        Tmp = state[i][0] ^ state[i][1] ^ state[i][2] ^ state[i][3];
        Tm  = state[i][0] ^ state[i][1]; Tm = xtime(Tm); state[i][0] ^= Tm ^ Tmp;
        Tm  = state[i][1] ^ state[i][2]; Tm = xtime(Tm); state[i][1] ^= Tm ^ Tmp;
        Tm  = state[i][2] ^ state[i][3]; Tm = xtime(Tm); state[i][2] ^= Tm ^ Tmp;
        Tm  = state[i][3] ^ t;            Tm = xtime(Tm); state[i][3] ^= Tm ^ Tmp;
    }
}
\end{lstlisting}
\vspace{-7pt}
  \end{subfigure}
  \caption{Hands-free C2HLSC AES Mix Columns}
  \label{fig:monobit}
\end{figure}

\begin{figure}[tb]
  \centering
  \begin{subfigure}[t]{.9\columnwidth}
    \begin{lstlisting}
HLS compatible code from the hands-free flow
    \end{lstlisting}
    \vspace{-7pt}

  \end{subfigure}
  \begin{subfigure}[t]{.9\columnwidth}
    \begin{lstlisting}[language=c]
#define getSBoxValue(num) (sbox[(num)])

// The SubBytes Function Substitutes the values in the
// state matrix with values in an S-box.
static void SubBytes(state_t state)
{
    uint8_t i, j;
    for (i = 0; i < 4; ++i)
    {
        for (j = 0; j < 4; ++j)
        {
            state[j][i] = getSBoxValue(state[j][i]);
        }
    }
}
\end{lstlisting}
\vspace{-7pt}
  \end{subfigure}
  \caption{Hands-free C2HLSC AES Sub Bytes}
  \label{fig:monobit}
\end{figure}

\subsection{Waveforms}
\subsection{Conversations} \label{app:conv}
Here we list the links to the conversations to go from C to HLS C.
\begin{itemize}
    \item \href{https://g.co/gemini/share/ba393c5de5a6}{Frequency Test: https://g.co/gemini/share/ba393c5de5a6}
    \item \href{https://g.co/gemini/share/659379d677c0}{Frequency Block Test: https://g.co/gemini/share/659379d677c0}
    \item \href{https://g.co/gemini/share/0f35a4d248e7}{Cumulative Sums Test: https://g.co/gemini/share/0f35a4d248e7}
    \item \href{https://g.co/gemini/share/92b68e7849fc}{QuickSort: https://g.co/gemini/share/92b68e7849fc}
    \item \href{https://g.co/gemini/share/92b68e7849fc}{AES 128: https://g.co/gemini/share/92b68e7849fc}
    \item \href{https://g.co/gemini/share/c1ac0cef56f2}{Overlapping Template Matching Test: https://g.co/gemini/share/c1ac0cef56f2}
    
\end{itemize}
\end{document}